\documentclass[iop,apjl]{emulateapj}

\setcitestyle{comma}

\usepackage{bm}

\shorttitle{Dark Matter HVCs Colliding with Magnetic Galaxy}
\shortauthors{Galyardt \and Shelton}

\begin{document}

\title{Collisions between Dark Matter Confined High Velocity Clouds
  and Magnetized Galactic Disks: The Smith Cloud}

\author{Jason Galyardt} 
\affil{Department of Physics and Astronomy, University of Georgia, Athens, GA  30602, USA}
\email{jeg@uga.edu}
\author{Robin L.~Shelton}
\affil{Department of Physics and Astronomy, University of Georgia, Athens, GA  30602, USA}
\email{rls@physast.uga.edu}

\date{\today}

\begin{abstract}

  The Galaxy's population of High Velocity Clouds (HVCs) may include a
  subpopulation that is confined by dark matter minihalos and falling
  toward the Galactic disk. We present the first magnetohydrodynamic
  simulational study of dark matter--dominated HVCs colliding with a
  weakly magnetized galactic disk. Our HVCs have baryonic masses of $5
  \times 10^6\,$M$_{\odot}$ and dark matter minihalo masses of 0, $3
  \times 10^8$, or $1 \times 10^9\,$M$_{\odot}$. They are modeled on
  the Smith Cloud, which is said to have collided with the disk 70~Myr
  ago. We find that, in all cases, the cloud's collision with the
  galactic disk creates a hole in the disk, completely disperses the
  cloud, and forms a bubble--shaped structure on the far side of the
  disk. In contrast, when present, the dark matter minihalo continues
  unimpeded along its trajectory. Later, as the minihalo passes
  through the bubble structure and galactic halo, it accretes up to
  $6.0 \times 10^5\,$M$_{\odot}$ in baryonic material, depending on
  the strengths of the magnetic field and minihalo gravity. These
  simulations suggest that if the Smith Cloud is associated with a
  dark matter minihalo and collided with the Galactic disk, the
  minihalo has accreted the observed gas. However, if the Smith Cloud
  is dark matter--free, it is on its first approach toward the
  disk. These simulations also suggest that the dark matter is most
  concentrated either at the head of the cloud or near the cloud,
  depending upon the strength of the magnetic field, a point that
  could inform indirect dark matter searches.

\end{abstract}


\section{Introduction}

The Milky Way has a population of gas clouds moving at high velocities
relative to the Local Standard of Rest, the so-called High Velocity
Clouds (HVCs).  Such clouds have also been observed in or near to
other large spiral galaxies \citep{Westmeier:2008,Miller:2009} and the
Local Group \citep{Adams:2013}.  They affect galaxy evolution by
delivering material that can be used for making stars
\citep{Putman:2006,Lehner:2011,Putman:2012,Fox:2014} and, upon impact
with the disk, by instigating star formation
\citep{Tenorio:1981,Lepine:1994,Izumi:2014}.

While the origins of HVCs are multiple and long debated \citep[see,
  for example,][]{Shapiro:1976,Gardiner:1996,Wakker:1996}, it is
likely that some HVCs are dark matter--confined and represent a
portion of the missing dark matter minihalo population
\citep{Blitz:1999,Braun:2000}. The extragalactic ultracompact HVCs in
the ALFALFA survey, with dynamical masses far exceeding their HI
masses, are excellent examples \citep{Adams:2013}.  These clouds have
$10^5$ to $10^6$~M$_\odot$ of HI mass and $10^7$ to $10^8$~M$_\odot$
of total mass, making them as massive as dwarf galaxies.  Even more
massive (in HI) is the Smith Cloud \citep{Smith:1963}, an HVC that is
claimed to have passed through the Milky Way's outer disk
\citep{Lockman:2008}, and, owing to its having survived that passage,
is suggested to be gravitationally confined by dark matter
\citep{Nichols:2009}.  Complex~H, an even more massive HVC presently
colliding with the outer disk of the Milky Way
\citep{Morras:1998,Izumi:2014}, has also been suspected of harboring
dark matter \citep{Simon:2006}.  Many dark matter--confined gaseous
clouds must have collided with a galaxy the size of the Milky Way
during its lifetime, calling attention to the question of how these
collisions affect both the clouds and the galaxy.

The dynamical evolution of the cloud--galaxy collision depends on
gravitational, magnetic, and hydrodynamic processes, often working
against each other. The collision of a pure baryonic cloud with a
non-magnetic galactic disk shocks and disrupts the cloud while
punching a hole in the disk \citep{Tenorio:1981,Tenorio:1986}.
However, a magnetized disk can act as a barrier to penetration of the
disk by a dark matter--free cloud \citep{Santillan:1999}. Conversely,
simulations in which the cloud has a dark matter cloud component and
the disk has no magnetic field result in a coherent gas cloud
appearing on the far side of the disk long after the collision
\citep{Nichols:2014}. The behavior of a dark matter--confined cloud
when impacting a magnetized galactic disk has not been reported
previously but is the topic of this work.

Using the three dimensional FLASH Eulerian multiphysics simulation
package \citep{Fryxell:2000}, we investigate the behavior of a dark
matter--confined HVC as it collides with a weakly magnetized galactic
disk similar in field geometry and gas density to that of the Milky
Way and compare with simulations that exclude one or both the
minihalo and the magnetic field.
We describe our simulation techniques and initial parameters in
Section~\ref{sec:methods}, discuss the results of individual
simulations in Section~\ref{sec:results}, and summarize and discuss
the ramifications for the Smith Cloud, infalling dark
matter--dominated clouds, and dark matter searches in
Section~\ref{sec:discussion}.

\section{Methodology and Simulations} \label{sec:methods}

Our cloud is patterned after the Smith Cloud, whose tip is currently
located at $(l, b) \approx (38^{\circ}.67,\,-13^{\circ}.41)$ and whose
heliocentric distance is $12.3 \pm 1.4\, \rm{kpc}$
\citep{Lockman:2008}. Its hydrogen mass exceeds $2 \times 10^6\,
\rm{M}_{\odot}$, evenly split between atomic
\citep{Lockman:2008} and ionized \citep{Hill:2009} hydrogen. The
observations by \citet{Hill:2013} of a significant line of sight
enhancement to the magnetic field around the Cloud indicate that it is
interacting with the local Galactic magnetic field. Following
\citet{Nichols:2014}, we model the progenitor cloud as a sphere of
0.5~kpc radius with an initial hydrogen number density of
0.4~cm$^{-3}$, which results in a total baryonic mass of $5.0 \times
10^{6}$~M$_{\odot}$. In our model, the cloud is initially in pressure
equilibrium with the ambient medium, resulting in an average interior
cloud temperature of $880\,\rm{K}$. The density and temperature
transition smoothly from those at the cloud center to those of the
ambient material at the cloud's periphery. In order to give the cloud
time to evolve dynamically in the gaseous halo environment prior to
the collision, we start the cloud 10~kpc above the midplane of the
Galactic disk, at a galactocentric radius of 13~kpc (the
galactocentric radius of its prior impact with the disk, as estimated
by \citet{Lockman:2008}), with a downward velocity of 200~km~s$^{-1}$.

For numerical tractability, we model the dark matter minihalo using
the Einasto density profile \citep{Einasto:1965,Merritt:2006}:
$\rho(r) = \rho_{e} \exp\{-d_{n}[(r/r_{e})^{1/n} - 1]\}$, where
$\rho_{e}$ defines the nominal density, $r_{e}$ defines the half-mass
radius for an infinite dark matter distribution, $n$ defines the
`shape' of the distribution, and $d_{n} \approx 3n - 1/3 + 0.0079/n$,
for $n \gtrsim 0.5$ \citep{Merritt:2006}. The models of
\citet{Nichols:2009} suggest that, in order for the Smith Cloud to survive its
passage through the Galactic disk, it must have a dark matter mass of
$2 \times 10^8$ to $1 \times 10^9$~M$_{\odot}$. In the interest of
exploring the sensitivity to the minihalo mass, we sample three minihalo
masses: 0, $3 \times 10^8$, and $1 \times 10^9$~M$_{\odot}$. For
consistency with the shape parameters used in \citet{Nichols:2009}, we
use common values of $r_{e} = 1.0\,\rm{kpc}$, and $n = 1 / 0.17$ in
all simulations that include a dark matter minihalo. For the smaller
mass minihalos, we use $\rho_{e} = 9.79 \times
10^{-3}\,\rm{M}_{\odot}\, pc^{-3}$; for the more massive minihalos, we
use $\rho_{e} = 3.27 \times 10^{-2}\,\rm{M}_{\odot}\, pc^{-3}$. We
model the dark matter in FLASH by approximately ten thousand massive,
collisionless particles. The particles' initial positions are
generated randomly from the appropriate Einasto density profile,
centered upon the baryonic cloud. In order to promote minihalo shape
stability, we generate a peculiar velocity for each particle according
to a Maxwell-Boltzmann distribution.  The total velocity of each
particle is set as the sum of the random peculiar velocity and the
minihalo group velocity of $200\,\rm{km}\,\rm{s}^{-1}$ toward the
disk.

We model the Galaxy's ISM as a cool disk \citep{Dehnen:1998} and hot
halo \citep{Miller:2013}, composed entirely of atomic hydrogen. We set
the gas density distribution, common to all our simulations, as the
sum of the disk and halo gas densities. The density at the midplane
within our simulation domain varies smoothly between 9.8 and 6.2
cm$^{-3}$ with increasing galactocentric radius. The density falls
smoothly with height, as set by the component models of
\citet{Dehnen:1998} and \citet{Miller:2013}. The temperature is 1100 K
at the Galactic midplane and it smoothly rises with height above the
plane until it reaches $1 \times 10^6$\,K at about 1.5\,kpc above the
midplane.

We model the distribution of mass within the Galaxy as the
superposition of the above--described gaseous components, a stellar
component that includes contributions from the bulge, thin and thick
disks \citep{McMillan:2011}, and halo \citep{Juric:2008}, and a dark
matter component whose distribution was adopted from
\citet{Navarro:1996} with parameters from \citet{McMillan:2011}. The
masses of these components sum to $1.6 \times 10^{12}\,
\rm{M}_{\odot}$.

We used this distribution of mass to calculate the Galaxy's
gravitational acceleration, $\bm{g}(R,z)$, as a function of position
in cylindrical polar coordinates. The galactic gravitational
acceleration in our simulations is thus self-consistent with the
observation--based mass distribution in our model.  In order to obtain
the gravitational acceleration at any spatial point, we first
calculate the 3D gravitational acceleration due to our mass model on
an $R$-$z$ grid with azimuthal symmetry and dimensions of $N \times
M$, resulting in $\bm{g}_{nm}(R_{n},z_{m}) = \sum_{i} g^{(i)}_{nm}
\hat{e}_{i}$, with the $\hat{e}_{i}$ representing the Cartesian unit
vectors. We then fit the grid of $g^{(i)}_{nm}$ with a separate
bivariate B-spline for each vector component, $i$. Our model's mass
distribution extends out to a galactocentric radius of $30\,\rm{kpc}$
and height of $\pm 200\,\rm{kpc}$.

For the runs that include a Galactic magnetic field, we implement the
coherent magnetic field component of \citet{Jansson:2012a} within the
simulation domain. This model includes spiral geometry in the disk,
transitioning to a toroidal geometry for the halo; it also includes an
`\textsf{X}'-shaped poloidal component. It is important to note that
the coherent field is roughly 15 to 20\% of the total field; the
remaining components of the \citet{Jansson:2012b} model are the
isotropic random and striated random fields. The latter two components
of the magnetic field are not included in the simulations presented
here due to numerical instabilities associated with turbulent flow of
the random field components under an ideal MHD evolution
framework. Thus, we characterize the magnetic field used in our
simulations as weak.

We use FLASH version 4.2 in three dimensions with a Cartesian
coordinate system. The procedural evolution of the runs within FLASH
varied slightly, employing either the unsplit hydrodynamic or the
unsplit staggered mesh ideal MHD solver, as appropriate. We use the
adiabatic ideal gas equation of state (EOS) with an adiabatic index
$\gamma$ of $5/3$. We neglect cooling. In order to account for both
the dynamic internal mass and the static external Galactic mass, we
implemented a new gravity solver in FLASH that combines the built--in
Barnes-Hut Tree Poisson solver for self-gravity with the interpolated
gravitational acceleration due to our Galactic mass model, yielding
the total gravitational acceleration. In our simulations, we ignored
the radial component of the Galaxy's gravitational acceleration,
setting $\bm{g}(R,z) = g_{z}(R,z) \hat{z}$, in order to minimize
radial motion and thus keep the simulation domain smaller and the
spatial resolution acceptable with the available computational
resources. The self-gravity component of the acceleration vector is
unrestricted in direction. While the baryonic mass in the simulation
domain contributes to both the self-gravity and the external gravity
source (via the Galactic mass model), the effect of this double
counting on the simulation dynamics is expected to be small. The
cloud's dark matter minihalo and the Galactic dark matter halo are the
dominant components of the self- and external gravity fields,
respectively, and are not affected by this double counting.

We performed a suite of simulations whose identification numbers, dark
matter mass, and presence of magnetic field are listed in
Table~\ref{table:runs}. 
Table~\ref{table:runs} also notes the baryonic
mass of the cloud, $m_{cloud}(t)$, at various stages of evolution. Our
method for calculating the baryonic cloud mass from the simulational
data was to integrate the gas density in three dimensions over a
sphere of $0.5\,\rm{kpc}$ radius.  The sphere was centered on the dark
matter minihalo barycenter for Runs 3--6. For completeness, we
performed our cloud mass calculation upon the runs without minihalos
and listed the results along with those of the other runs in
Table~\ref{table:runs}. For these runs (Runs 1 \& 2), the sphere was
centered either on the baryonic cloud, or, if no cloud was obvious, on
the lowest $z$ point of the bubble's shell (thus the point most
similar to that in Runs 3--6) for pre-collision and post-collision
epochs, respectively. We did not subtract any ambient baryonic mass
that the sphere contained. The dark matter--free runs develop at a
slower rate after the collision than do the runs with minihalos,
requiring an additional 7~Myr to reach the same level of
maturity. Therefore, the epoch labeled $t_{2}$ in
Table~\ref{table:runs} corresponds to a simulation time of 52~Myr for
Runs 1 \& 2 and 45~Myr for Runs 3--6. Epochs labeled $t_{1}$ and
$t_{3}$ correspond to 35 and 75~Myr, respectively, for all runs.

\begin{deluxetable}{crcccc} 
  \tablewidth{\columnwidth}
  \tablecolumns{6}
  \tablecaption{Simulation Runs: Physical Effects and Baryonic Cloud Masses}
  \tablehead{            
    \colhead{Run}    & \colhead{$m_{DM}\,[\rm{M}_{\odot}]$} & \colhead{\textbf{\textit{B}}-field} & \multicolumn{3}{c}{$m_{cloud}(t)\,[\rm{M}_{\odot}]$} \\[0.1cm]
    \cline{4-6} \\[-0.2cm]
    &  &  & $t_{1}$ & $t_{2}$ & $t_{3}$ \\[-0.3cm] 
  }
  \startdata %
  1 & $0.0$              & No & $3.2 \times 10^{6}$ & $3.9 \times 10^{5}$ & $2.1 \times 10^{5}$ \\ 
  2 & $0.0$              & Yes & $3.2 \times 10^{6}$ & $5.2 \times 10^{5}$ & $3.7 \times 10^{5}$ \\ 
  3 & $3.0 \times 10^{8}$ & No & $4.1 \times 10^{6}$ & $2.2 \times 10^{5}$ & $9.4 \times 10^{4}$ \\ 
  4 & $3.0 \times 10^{8}$ & Yes & $4.2 \times 10^{6}$ & $1.4 \times 10^{5}$ & $2.5 \times 10^{4}$ \\ 
  5 & $1.0 \times 10^{9}$ & No & $4.1 \times 10^{6}$ & $6.2 \times 10^{5}$ & $7.9 \times 10^{5}$ \\ 
  6 & $1.0 \times 10^{9}$ & Yes & $4.1 \times 10^{6}$ & $4.6 \times 10^{5}$ & $6.0 \times 10^{5}$    
  \enddata
  \tablecomments{%
  For all runs, the initial baryonic cloud
  mass, $m_{cloud}(t_{0})$, was $5.0\,\times\,10^{6}\,\rm{M}_{\odot}$. See the text for the
  method used to calculate $m_{cloud}(t)$ for later times. The
  sampling times for baryonic mass were as follows: $t_{1} =
  35\,\rm{Myr}$ for all runs; $t_{2} = 52\,\rm{Myr}$ for Runs 1 \& 2
  and $t_{2} = 45\,\rm{Myr}$ for Runs 3--6; $t_{3} = 75\,\rm{Myr}$ for
  all runs. The second epoch, $t_{2}$, differs in sampling time according to
  dark matter content: after the collision, the dark matter--free runs
  require an additional 7~Myr to develop to the same level of maturity
  as those with minihalos. }
  \label{table:runs}
\end{deluxetable}

\begin{figure}
  \plotone{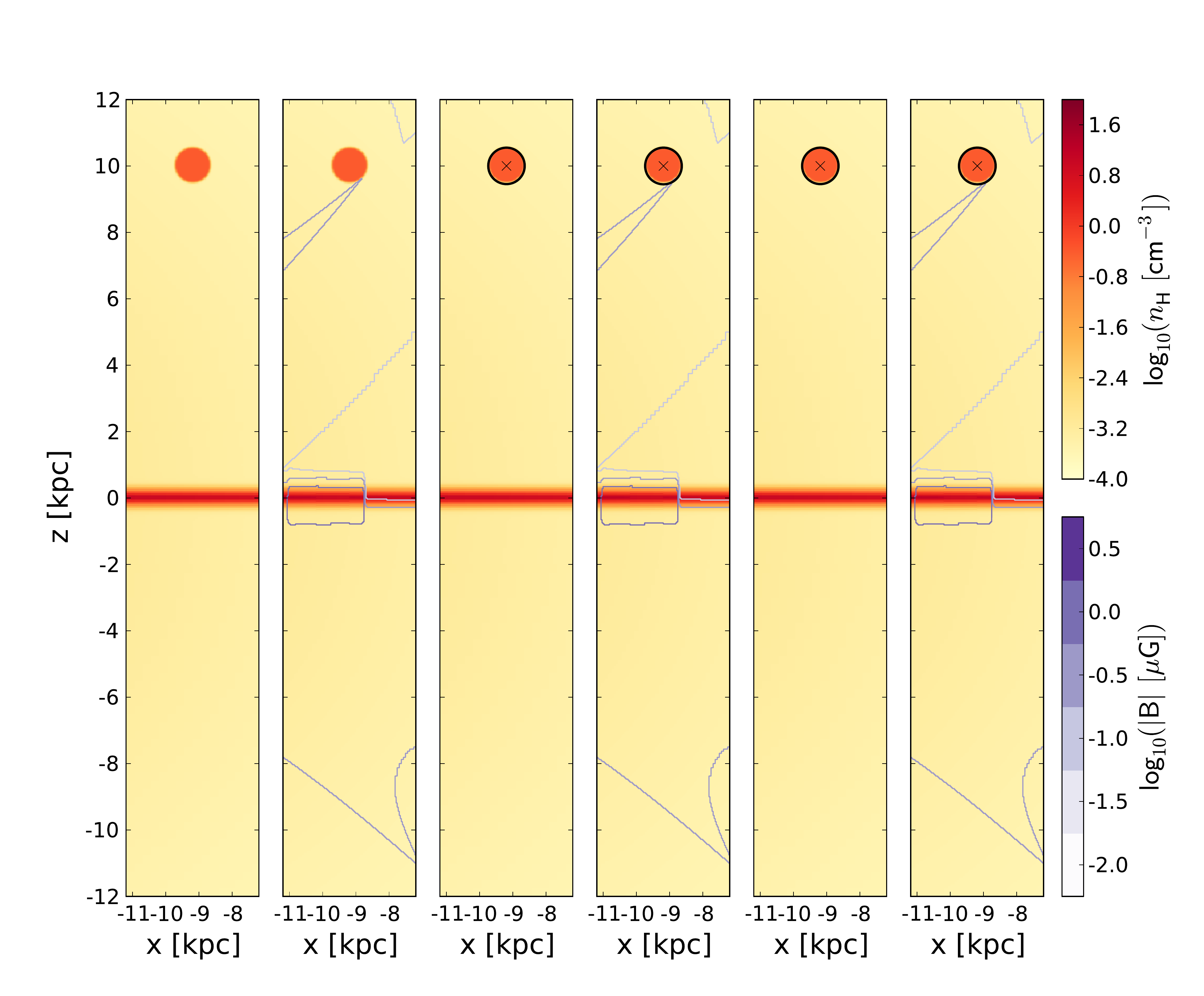}
  \caption{Initial conditions for each simulation along slices
    through the $y=0$ plane. $\log_{10}$ of hydrogen number density is
    represented by the continuous color scale. $\log_{10}$ of magnetic
    field strength is represented by contours. The dark matter
    barycenter is marked with an `\textsf{X}' and a circle at the
    half-mass radius. From left to right, the panels show Runs 1--6,
    as in Table~\ref{table:runs}.}
  \label{fig:t0Myr}
\end{figure}

Figure~\ref{fig:t0Myr} shows the initial conditions for all six runs
in slices through the $y = 0$ plane. For computational efficiency, we
restrict the simulation domain to a $4\, \rm{kpc} \times 4\, \rm{kpc}
\times 24\, \rm{kpc}$ region of the Galaxy. The Galactic magnetic
field model of \citet{Jansson:2012a} has azimuthal variations in
strength and direction of the field. We chose the galactocentric
azimuthal angle to roughly agree with that of the point of impact 70
Myr ago shown in \citet{Nichols:2009}, Figure 3. The minihalos are
necessarily constrained by the size of the domain and fill spheres of
$2\, \rm{kpc}$ radius (the domain half-width). In
Figure~\ref{fig:t0Myr}, the minihalo barycenters are indicated by the
`\textsf{X}' markers, and their simulation domain half-mass radii
(distinct from $r_{e}$) are shown as black circles.

\section{Results} \label{sec:results}

\begin{figure}
  \plotone{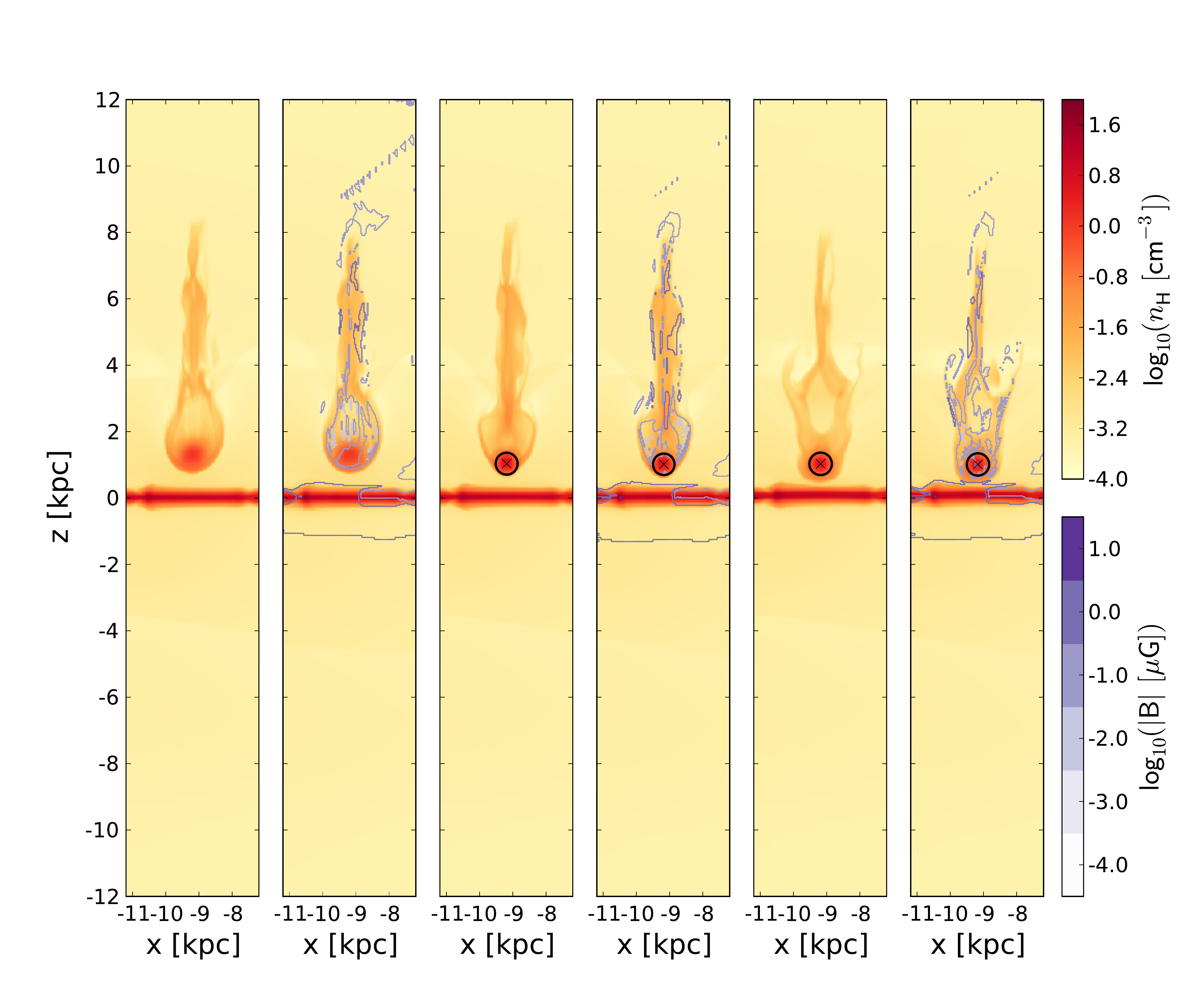}
  \caption{Same as Figure~\ref{fig:t0Myr}, but for epoch $t_{1}$,
    equivalent to $t = 35 \rm{Myr}$ for all runs.}
  \label{fig:t35Myr}
\end{figure}

All six clouds behave similarly as they fall through the halo. This
can be seen in Figure~\ref{fig:t35Myr} which shows the clouds at epoch
$t_{1}$ (35~Myr), just before impact.  As they pass through the halo,
the clouds lose material due to ram pressure stripping.  By epoch
$t_{1}$ the clouds with minihalos have lost 16--18\% of their original
baryonic mass, while those without have lost $\sim$36\% (see
Table~\ref{table:runs}). During this stage, the effect of the weak
magnetic field is merely to become swept up and wrapped around the
cloud.  This is similar to the behavior in the 2D MHD simulations of
\citet{Konz:2002}.


\begin{figure}
  \plotone{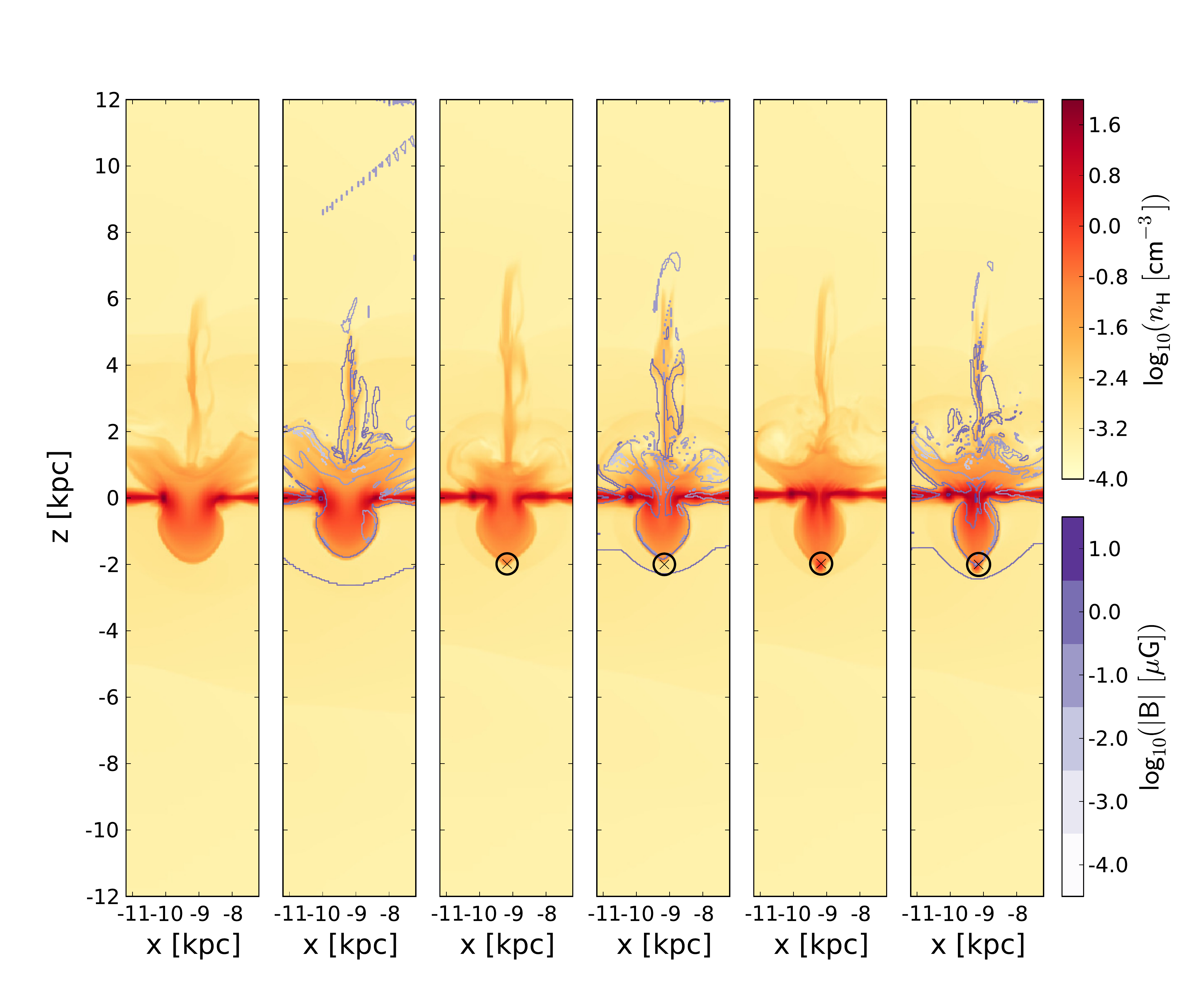}
  \caption{Same as Figure~\ref{fig:t0Myr}, but for epoch $t_{2}$,
    equivalent to $t = 52\, \rm{Myr}$ for Runs 1 \& 2 and $t = 45\,
    \rm{Myr}$ for Runs 3--6.}
  \label{fig:t45Myr}
\end{figure}

Each of the clouds approaches the disk with a similar velocity
($\sim$300~km~s$^{-1}$).  However, the dark matter confined clouds are
more compact and contain more baryonic mass and so carry larger
kinetic energy densities than the dark matter--free clouds.  The
collision of the cloud with the Galactic disk occurs at 36\,Myr for
those runs with a minihalo, and 37\,Myr for the dark matter--free
runs.  At the moment of collision, the minihalo separates from the
cloud gas and continues, unhindered, through the Galactic disk.
During the collision, the gas cloud ceases to exist as it mixes with
and transfers momentum to the disk gas.  This results in a hole in the
disk and a long--lived, bubble--shaped structure filled with
moderately dense ($\sim$0.1 to $\sim$3~H~cm$^{-3}$), cool gas
($\sim$1000~K) below the disk, as shown in Figure~\ref{fig:t45Myr} at
52~Myr for Runs 1 \& 2 and 45~Myr for Runs 3--6 (epoch $t_{2}$). These
structures are composed of a mixture of gas from both the cloud and
the disk, with the majority of their mass originating from the
disk. \citet{Tenorio:1986} found similar structures in 2D hydrodynamic
simulations of dark matter--free clouds colliding with the Galactic
thin disk.

Over time, each minihalo gravitationally accretes gas from its
surroundings. By epoch $t_{2}$, the minihalos have trapped an amount
of baryonic mass equivalent to 2.8 to 12\% of the original baryonic
cloud mass.  See Table~\ref{table:runs}.

\begin{figure}
  \plotone{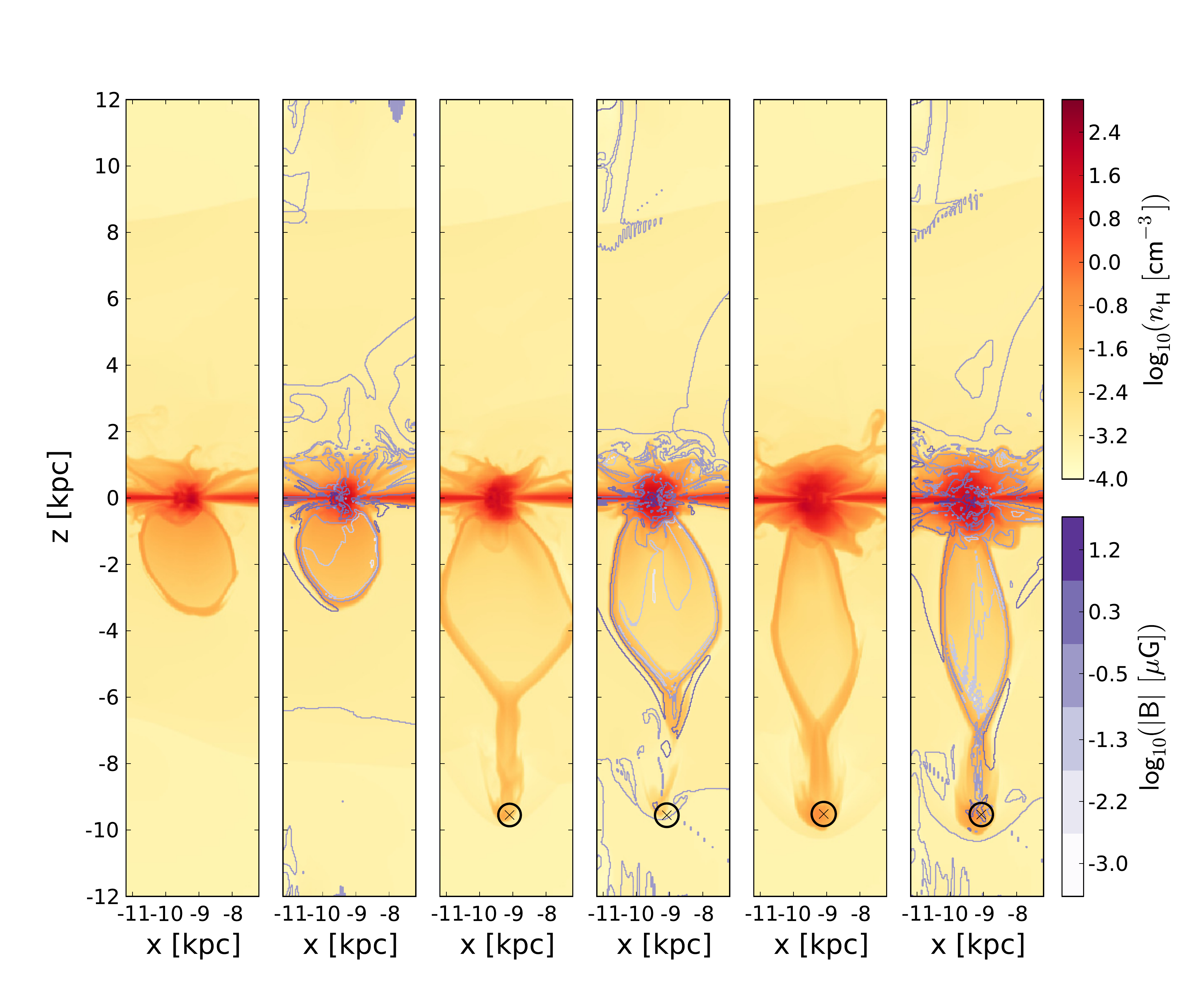}
  \caption{Same as Figure~\ref{fig:t0Myr}, but for epoch $t_{3}$,
    equivalent to $t = 75 \rm{Myr}$ for all runs.}
  \label{fig:t75Myr}
\end{figure}

For runs 3--6, the magnetohydrodynamic and dark matter gravitational
forces continue to influence the evolution of the gas density
distribution long after the collision (Figure~\ref{fig:t45Myr} to
Figure~\ref{fig:t75Myr}).  The minihalo's gravity continues to tug on
the gas at the bottom of this bubble--shaped structure while
magnetohydrodynamic forces slow its horizontal expansion. Although the
gravitational tug elongates this structure in simulations 3--6
relative to those in Runs 1 \& 2, it cannot do so
indefinitely. Eventually it pulls off a small, comet-shaped
fragment. In Run 4, the fragment completely separates from the
bubble--shaped structure, but in Runs 3, 5, and 6, a drip-line of gas
connects the fragment to the bubble--shaped structure. In Runs 3 \& 4,
where the minihalo mass is $3 \times 10^8\, \rm{M}_{\odot}$, the 
newly accreted baryonic mass is very small, only 1.9 and 0.50\% of the original
baryonic mass, respectively. In contrast, the $3 \times$ more massive
dark matter minihalos in Runs 5 and 6 allow them to accrete gas from
the surrounding Galactic halo, bringing the mass of their freshly
formed clouds to 16 and 12\%, respectively, of their original baryonic
masses by epoch $t_{3}$ (75~Myr).

\section{Summary and Discussion} \label{sec:discussion}

It has been suggested that a dark matter minihalo ``shepherded'' the
Smith Cloud through the outer disk, that the minihalo then
ballistically traveled through the halo, dragging the gas cloud along
with it, and that they are now poised to fall into the disk again.  In
order to examine the passage of a dark matter enshrouded Smith
Cloud-sized gas cloud through the outer portion of a magnetized
galactic disk, we performed a series of simulations, some with a weak
galactic magnetic field and some with a dark matter minihalo.  In all
cases, the cloud's collision with the disk punches a hole in the disk,
obliterates the cloud, and produces a large bubble--shaped
distribution of cool gas below the disk.  When a dark matter minihalo
is present, it gathers material from the bubble into a comet--shaped
density enhancement that travels with the minihalo.  In addition, if
the minihalo is extremely massive, it later accretes some gas from the
Galactic halo, and the head of the cloud likely remains co--located
with the minihalo. It is thus a prime target for dark matter signal
searches. It should be noted that the simulation with a low mass
minihalo and a galactic magnetic field (Run 4) shows indications that
the gas currently confined by the minihalo may be stripped out by the
galactic magnetic field on time scales longer than we simulated,
resulting in the minihalo leading the cloud. However, in none of our
cases is the final mass of baryonic material accompanying the minihalo
more than 16\% of the primordial cloud mass or 40\% of the observed
Smith Cloud mass.

The presence of even a weak Galactic magnetic field reduces the amount
of gas that accompanies the minihalo, irrespective of minihalo mass.
It should be noted, however, that only the coherent component of the
Galactic magnetic field is modeled in this work. If the random
components of the \citet{Jansson:2012b} model were to be included as
well (for fixed minihalo mass), then the net strength of the magnetic
field in the disk and halo would have been $\sim$6 times greater and
would have resulted in a smaller final cloud mass.

Under the premise that the Smith Cloud previously passed through the
Galactic disk accompanied by a dark matter minihalo, our simulations
indicate that the observed gas comprising the Smith Cloud must have
been accreted onto the minihalo since its disk passage. This implies
that the metallicity of the Smith Cloud should be similar to that of
the Galactic ISM along its trajectory, rather than primordial in
nature. Our simulations favor the larger mass minihalo under this
premise, as it is likely that the Galactic magnetic field would
prevent the lower mass minihalo from accumulating the amount of gas
observed.

On the other hand, if the premise is not correct and the Smith Cloud
was not accompanied by a dark matter minihalo, then it would not have
survived passage through the Galactic disk, irrespective of the
presence of a magnetic field.  In that case, the Smith Cloud must
currently be on its first descent into the disk.

From the Galactic point of view, the remnant bubbles of HVC--disk
collisions grow to heights of 3 to 7~kpc. Based in extended simulation
times on Runs 1, 3, and 5, these structures should survive for $\sim$50~Myr (for
HVCs without dark matter) to $\sim$60~Myr (for HVCs with dark
matter). If HVC material were to collide with a weakly magnetized
galactic disk at the rate of 1~M$_{\odot}$\,yr$^{-1}$
\citep{Shull:2009,Lehner:2011} in $5 \times 10^6$M$_{\odot}$ clouds as
in our simulations, we would expect $\sim$10 to 12 extant bubbles at
any given time, for purely dark matter--free and purely dark
matter--rich HVC--disk collisions, respectively. The number of HVC
collision--driven bubbles would be affected by the actual Galactic
magnetic field strength (a stronger field would result in fewer
bubbles at any given time), the number of HVCs reaching the disk, the
baryonic mass and size of impacting HVCs, and the proportion of
gas--bearing minihalos on collisional orbits.

Our simulated HVC collision--driven bubble structures are reminiscent
of the supergiant shells (SGSs) observed in the Milky Way
\citep{Heiles:1984} and elsewhere (e.g., LMC: \citealt{Book:2008};
M31: \citealt{Brinks:1986}; IC 2574: \citealt{Walter:1999}).  Young
SGSs are likely to contain hot gas \citep{McCray:1987} and have
enhanced metallicities due to supernova activity. Evolved SGSs are
likely to have cool interiors due to metal line emission
\citep{MacLow:2000} or blowouts \citep{McCray:1987}. In contrast, we
predict that HVC collision--driven structures would have cool
interiors and galactic or sub--galactic metallicities.  In principle,
one could perform a search for HVC collision--driven shell structures
within the population of Galactic or extragalactic SGSs. Using the
discriminants outlined above, such a search may allow one to identify
candidate structures and constrain the rate of infalling gas--bearing
minihalos.

Our simulations show that the majority of the infalling HVC gas will
merge into the Galactic disk. Although the collision blows an
enormous bubble, most of the bubble material will eventually fall back
into the disk. Only a small amount of HVC plus Galactic material will
be carried away with the dark matter minihalo. HVC-disk collisions are
thus efficient mechanisms for contributing gas to the Galactic disk,
supporting future star formation.

\acknowledgments We would like to acknowledge J.\ Bland-Hawthorn,
G.\ Farrar, A.\ Hill, and F.~J.\ Lockman for useful discussions and
the anonymous referee for insightful comments. This study was
supported in part by resources and technical expertise from the
Georgia Advanced Computing Resource Center, a partnership between the
University of Georgia's Office of the Vice President for Research and
Office of the Vice President for Information Technology. This work was
supported by NASA grant 10-21-RR185-451.

\end{document}